\documentclass[aps,prc,twocolumn,superscriptaddress]{revtex4-2}
\usepackage{xcolor}
\usepackage{graphicx}
\usepackage{subfig}    
\usepackage{overpic}
\usepackage{caption}
\usepackage{booktabs} 

\begin{document}


\title{Pseudorapidity dependence of charged particles production in non-single diffractive $pp$ collisions in the PACIAE 4.0 model}


\author{Zhen Xie}
\affiliation{School of Physics and Information Technology, Shaanxi Normal University, Xi'an 710119, China}


\author{An-Ke Lei}
\affiliation{School of Physics and Electronic Science, Guizhou Normal University, Guiyang, 550025, China}
\affiliation{Key Laboratory of Quark and Lepton Physics (MOE) and Institute of Particle Physics, Central China Normal University, Wuhan 430079, China}

\author{Hua Zheng}
\email[]{zhengh@snnu.edu.cn}

\author{Wen-Chao Zhang}
\affiliation{School of Physics and Information Technology, Shaanxi Normal University, Xi'an 710119, China}

\author{Dai-Mei Zhou}
\email[]{zhoudm@mail.ccnu.edu.cn}
\affiliation{Key Laboratory of Quark and Lepton Physics (MOE) and Institute of Particle Physics, Central China Normal University, Wuhan 430079, China}

\author{Zhi-Lei She}
\affiliation{School of Mathematical and Physical Sciences, Wuhan Textile University, Wuhan 430200, China}

\author{Yu-Liang Yan}
\affiliation{China Institute of Atomic Energy, P. O. Box 275 (10), Beijing 102413, China}

\author{Ben-Hao Sa}
\email[]{sabhliuym35@qq.com}
\affiliation{Key Laboratory of Quark and Lepton Physics (MOE) and Institute of Particle Physics, Central China Normal University, Wuhan 430079, China}
\affiliation{China Institute of Atomic Energy, P. O. Box 275 (10), Beijing 102413, China}


\date{\today}

\begin{abstract}
Studying experimental observables is a key benchmark for validating theoretical models in high energy physics. In this work, we employ the PACIAE 4.0 model to simulate non-single diffractive proton-proton ($pp$) collisions at center-of-mass energies of 0.9, 2.36, and 7 TeV, comparing the results with Compact Muon Solenoid (CMS) experimental data on charged-particle pseudorapidity densities and transverse momentum spectra across different pseudorapidity bins, respectively. Our results show good agreement with the CMS data, particularly only using  a single set of parameters for all collision energies. This demonstrates that the PACIAE 4.0 model can serve as a reliable tool for systematically studying the physics of NSD $pp$ collisions.
\end{abstract}


\maketitle

\section{INTRODUCTION}{\label{intro}}
Proton-proton ($pp$) collisions at ultra-relativistic energies have emerged as a complementary approach to studying the properties of quark-gluon plasma (QGP), alongside relativistic nucleus-nucleus ($AA$) collisions. In particular, high-multiplicity $pp$ collision events exhibit collective behavior resembling small QGP droplets \cite{ALICE:2016fzo,CMS:2016fnw,STAR:2005gfr,PHOBOS:2004zne}. The pseudorapidity density distributions ($dN_{ch}/d\eta$) and transverse momentum ($p_{\rm{T}}$) spectra of charged particles are fundamental experimental observables, carrying information about system dynamics at freeze-out and even earlier stages \cite{PHOBOS:2010eyu,STAR:2006xud,STAR:2006nmo,ALICE:2013jfw,ALICE:2015olq,ALICE:2019hno,ALICE:2014juv,ALICE:2015ial,ALICE:2015qqj,ALICE:2010cin,ALICE:2011gmo,ALICE:2022kol}. Moreover, these measurements provide insights into both soft and hard Quantum Chromodynamics (QCD) processes, as well as particle production mechanisms. Thus, studying these observables is crucial for understanding QGP properties \cite{Werner:2024ntd,Waqas:2022omn,Ajaz:2022mga,Tao:2023kcu,Yang:2022fcj,Zhang:2025pqu,STAR:2017sal,Toia:2011nzq,Deppman:2019yno,Shi:2024pyz,Tao:2023kcu,Tao:2022tcw,Zhu:2022dlc,Zhu:2022bpe,Tao:2020uzw,Zhu:2021fbs,Zheng:2015mhz,Wang:2022det,Zhu:2018nev,Gao:2017yas,Zheng:2015gaa,Zheng:2015tua,Wong:2015mba,Rath:2019cpe,Xu:2017akx,Yang:2022fcj,Zhao:2020wcd,Zhao:2021vmu}. However, high energy $pp$ collisions are highly complex dynamical processes. Experimental measurements alone cannot fully characterize the underlying physics, necessitating theoretical models, such as PYTHIA~\cite{Sjostrand:2006za,Bierlich:2022pfr}, 
HERWIG \cite{Corcella:2000bw}, HIJING \cite{Wang:1991hta}, 
AMPT \cite{Lin:2004en}, SMASH \cite{PhysRevC.94.054905}, PACIAE \cite{Sa:2011ye,Lei:2023srp,Lei:2024kam}, etc., to describe the collision evolution across different stages. The measured observables not only validate theoretical models but also impose stringent constraints on their parameter space \cite{Skands:2010ak,Skands:2014pea,CMS:2015wcf}.

In our previous work \cite{Xie:2025vnh}, we showed that the PACIAE 4.0 model can reproduce well the experimental data for the pseudorapidity density distributions of charged particles and the transverse momentum spectra of identified particles at midrapidity in inelastic (INEL) $pp$ collisions at various collision energies. This provided a valuable resource for both experimentalists and theorists since $pp$ collisions serve as the baseline for heavy-ion collisions. In 2010, the Compact Muon Solenoid (CMS) Collaboration at the Large Hadron Collider (LHC) published results on the pseudorapidity density distributions and the pseudorapidity dependence of the transverse momentum spectra for the charged particles produced in non-single diffractive (NSD) $pp$ collisions at centre-of-mass energies ($\sqrt{s}$) of 0.9, 2.36 and 7 TeV \cite{CMS:2010wcx,CMS:2010tjh}. These data offer a valuable opportunity to further validate the capabilities of the PACIAE 4.0 model in simulating NSD $pp$ collisions. Therefore, we simulate the pseudorapidity dependence of the transverse momentum spectra and the pseudorapidity density distributions for charged particles in the NSD $pp$ collisions at $\sqrt{s}=$0.9, 2.36, and 7 TeV using the PACIAE 4.0 model. Additionally, we investigate and discuss the differences in model parameters for NSD and INEL $pp$ collisions. 

The paper is organized as follows. In Sec.~\ref{model}, the PACIAE 4.0 model is briefly introduced. In Sec.~\ref{res}, we present the PACIAE 4.0 simulation results for both the pseudorapidity density distributions and the pseudorapidity dependence of the transverse momentum spectra of charged particles in NSD $pp$ collisions at $\sqrt{s}=$0.9, 2.36 and 7 TeV, including comparisons with experimental data. A brief summary is drawn in Sec.~\ref{con}. 
 
\section{The PACIAE 4.0 model}\label{model}
PACIAE \cite{Sa:2011ye,Lei:2023srp,Lei:2024kam} is a phenomenological parton and hadron cascade model based on PYTHIA \cite{Sjostrand:2006za, Bierlich:2022pfr}. It is designed to describe high-energy collisions involving lepton-lepton, lepton-hadron, lepton-nucleus, hadron-hadron, hadron-nucleus, and nucleus-nucleus interactions. Same as its earlier versions \cite{Sa:2011ye,Lei:2023srp}, the latest released version of PACIAE 4.0 \cite{Lei:2024kam} comprises four stages: the initial state, parton cascade (partonic rescattering), hadronization, and hadron cascade (hadronic rescattering). 

\begin{figure}[h]
\centering
\includegraphics[width=1\linewidth]{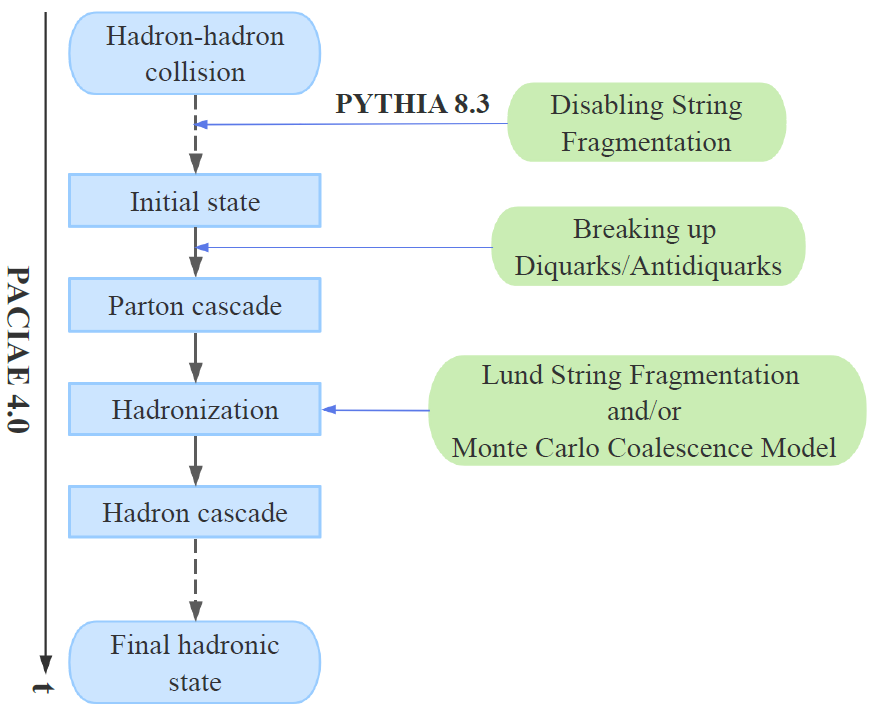}
\caption{The program flow for a $pp$ collision in PACIAE 4.0 model.}
\label{fig:flow}
\end{figure}

For a $pp$ collision in the PACIAE 4.0 model, the initial state is first generated by PYTHIA 8.3 with the string fragmentation temporarily switched off. The strings are then broken down, and the diquarks split randomly resulting in an initial partonic state. Partonic rescattering is subsequently considered through $2\rightarrow 2$ parton-parton scatterings using leading-order perturbative quantum chromodynamics (pQCD) cross sections \cite{Combridge:1977dm}. In the sequential hadronization process, partons are hadronized through the Lund string fragmentation approach (used in this work) or via the coalescence model. Finally, hadronic rescattering is executed until kinetic freeze-out. A more comprehensive description of the PACIAE 4.0 model can be found in our recent works \cite{Lei:2024kam, Xie:2025vnh}. The program flow for a $pp$ collision in PACIAE 4.0 is shown in Fig.~\ref{fig:flow}.  

There are three adjustable parameters in the $pp$ collision simulation in PACIAE 4.0: The $K$ factor (adj1(10) in the program) is a multiplicative factor of the hard scattering cross sections \cite{Combridge:1977dm,Field:1989uq}, i.e.,

\begin{equation}
    \frac{d\sigma}{dt}(ab\rightarrow cd;s,t)= K\frac{\pi\alpha_s^2}{s^2}|\bar{M}(ab\rightarrow cd)|^2,
\end{equation}
where $\alpha_s$ is the strong coupling constant. $s$ and $t$ are the Mandelstam invariants in the kinematics of the $ab\rightarrow cd$ parton-parton process. The other two parameters are $a$ and $b$ in the LUND fragmentation probability function \cite{Sjostrand:2006za,Andersson:1983ia}: 
\begin{equation}
    f(z) \propto (1/z)(1-z)^a\exp(-b m_T^2 / z), \label{lf}
\end{equation}
which is closely related to the particle production in the simulation. In Eq. (\ref{lf}), $z$ denotes the energy fraction taken away by a hadron fragmented from a high energy parton; $m_T=\sqrt{m^2+p_T^2}$ is the transverse mass of the hadron.

\begin{figure*}[ht]
   \centering
    \subfloat{
	\begin{overpic}[width=0.45\linewidth]{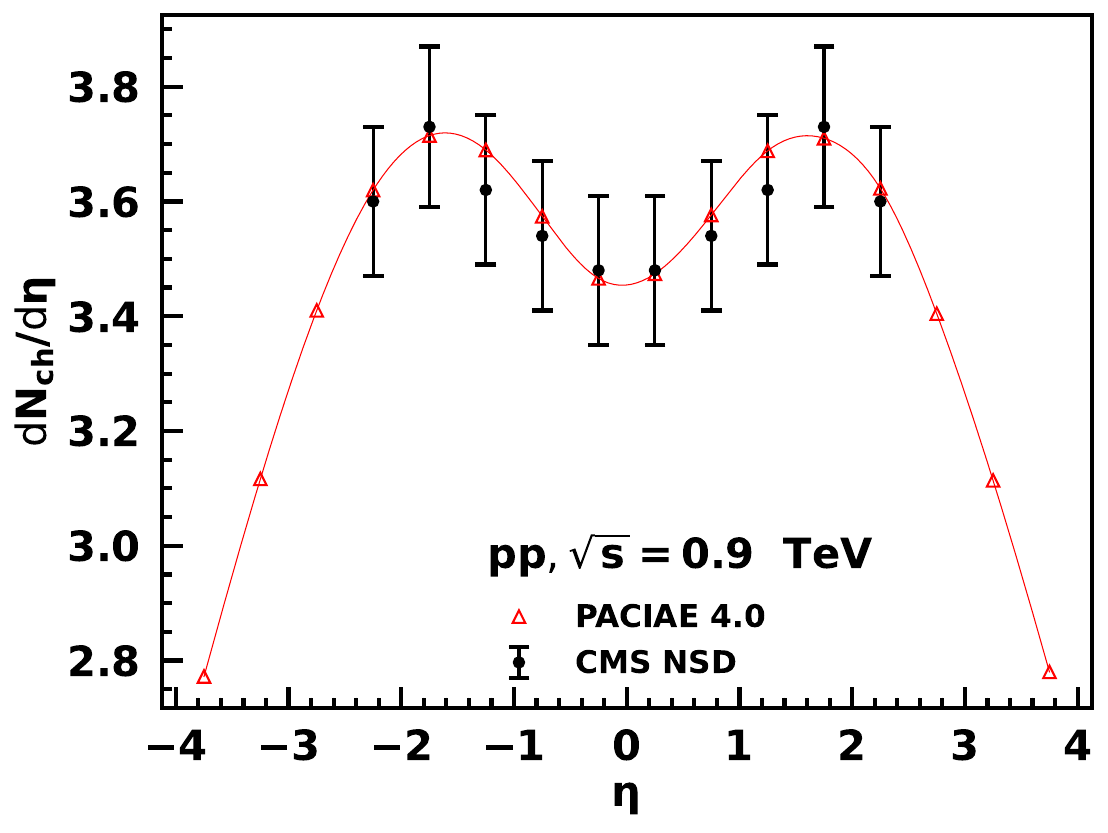}
    \end{overpic}
	}
    \hfill
     \subfloat{
	\begin{overpic}[width=0.45\linewidth]{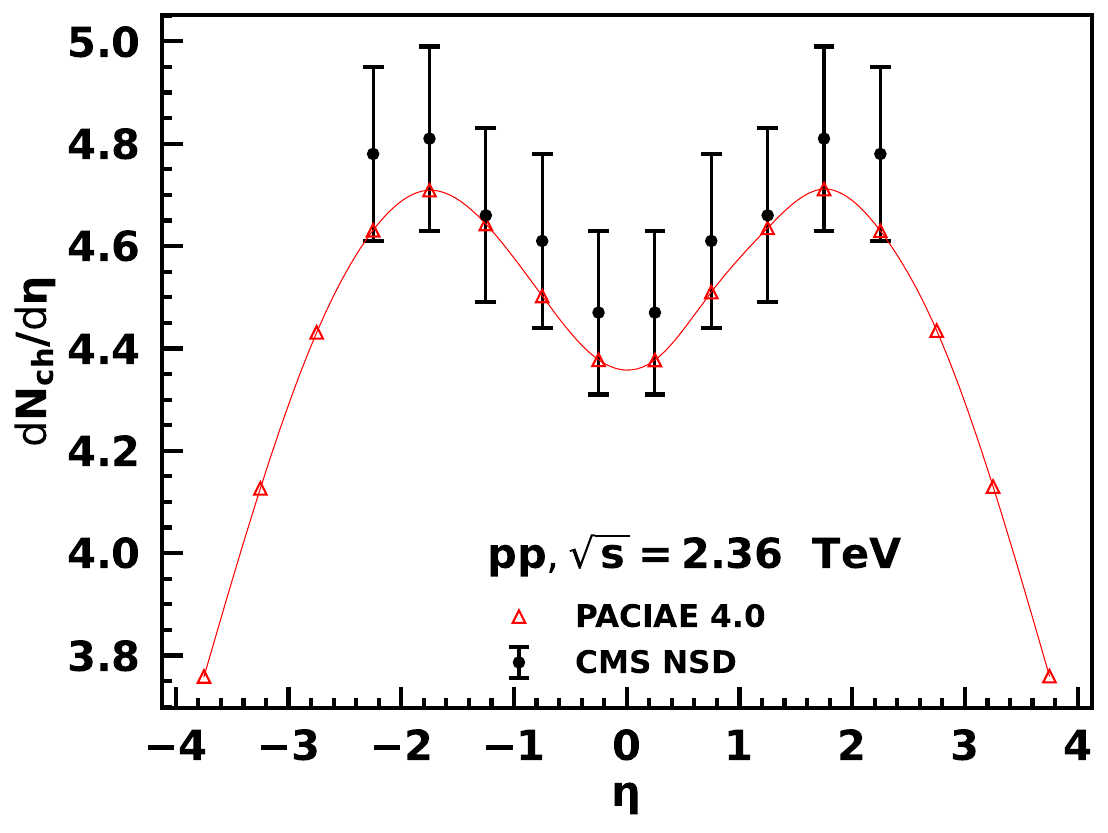}
    \end{overpic}
	}
    \hfill
    \subfloat{
	\begin{overpic}[width=0.45\linewidth]{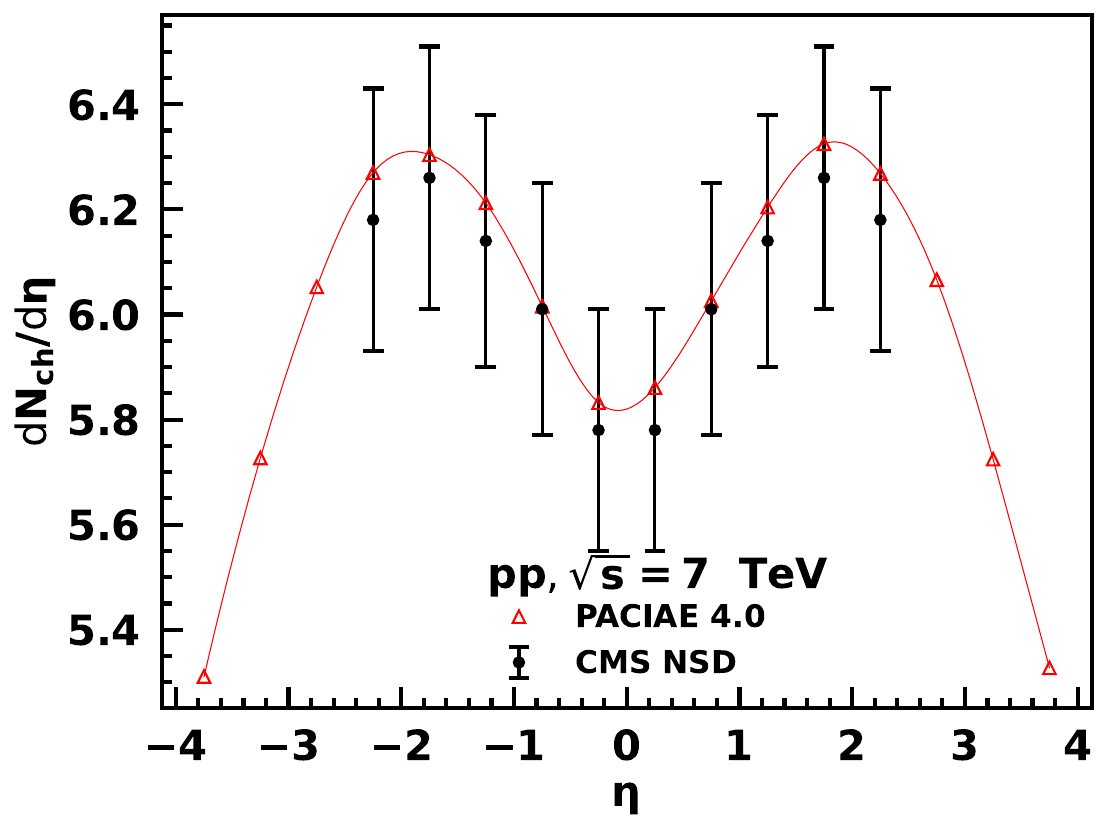}
    \end{overpic}
	}
   \caption{Pseudorapidity density distributions of charged particles produced in NSD $pp$ collisions at $\sqrt{s}=$ 0.9, 2.36, and 7 TeV from PACIAE 4.0 model simulations are compared with CMS data \cite{CMS:2010wcx,CMS:2010tjh}.}
  \label{fig2}
\end{figure*}

\section{RESULTS AND DISCUSSION }\label{res}

In this study, we ultilize CMS data for both pseudorapidity density distributions and pseudorapidity dependence of the transverse momentum spectra of charged particles in NSD $pp$ collisions at three different collision energies, comparing them with PACIAE 4.0 simulations. The three adjustable parameters in our model with values of:
 
\begin{itemize}
\item $K=0.8$: It is same as in Ref. \cite{Xie:2025vnh}. This indicates that we require the same hard-scattering cross sections in NSD $pp$ collisions as those in INEL $pp$ collisions to reproduce the corresponding experimental data at the same collision energy in the PACIAE 4.0 model.  
\item $a=0.68$: Consistent with Ref. \cite{Xie:2025vnh}, this parameter remains unchanged.   
\item $b=0.8$: It is a free parameter to optimize the agreement between the simulation results from PACIAE 4.0 and the experimental data for pseudorapidity density as well as the pseudorapidity dependence of transverse momentum distributions of charged particles simultaneously at a given collision energy. Following the parameter adjustment strategy of Ref. \cite{Xie:2025vnh}, we surprisingly found $b=0.8$ works consistently across all the three collision energies studied.
\end{itemize}
All other input parameters retain their default PACIAE 4.0 values.

\begin{figure*}[ht]
   \centering
    \subfloat{
	\begin{overpic}[width=0.48\linewidth]{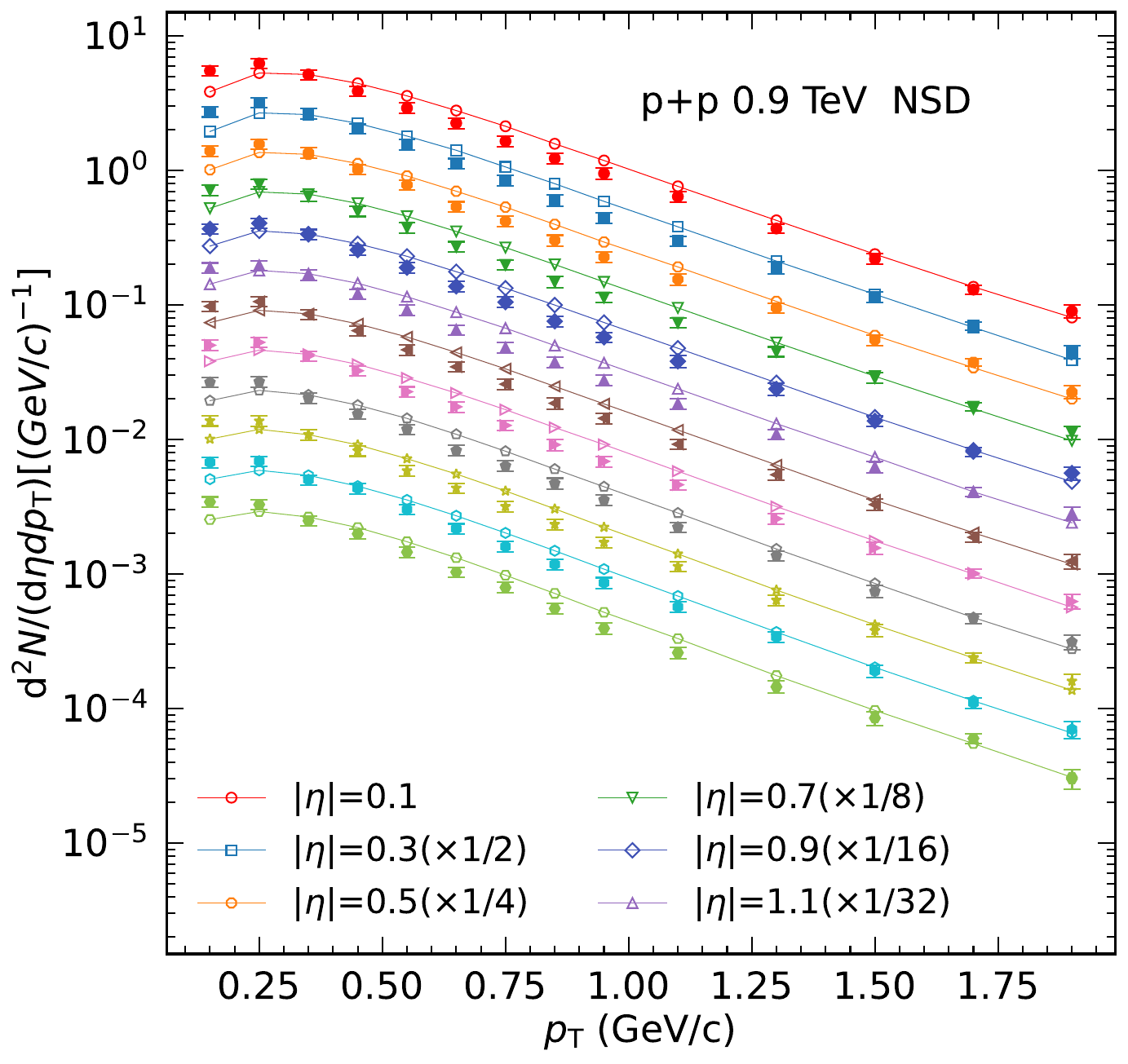}
    \end{overpic}
	}
    \hfill
     \subfloat{
	\begin{overpic}[width=0.48\linewidth]{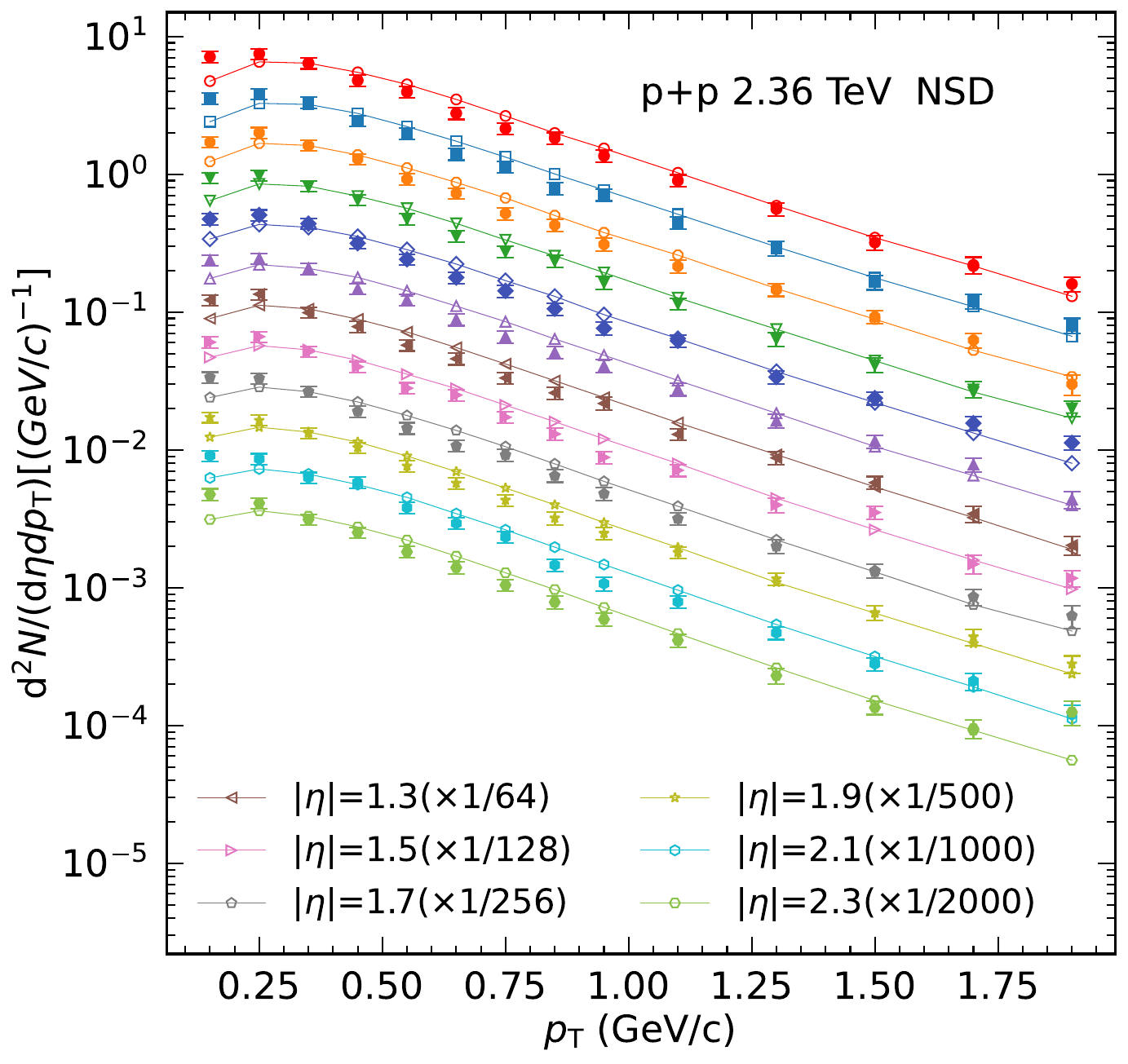}
    \end{overpic}
	}
    \hfill
    \subfloat{
	\begin{overpic}[width=0.48\linewidth]{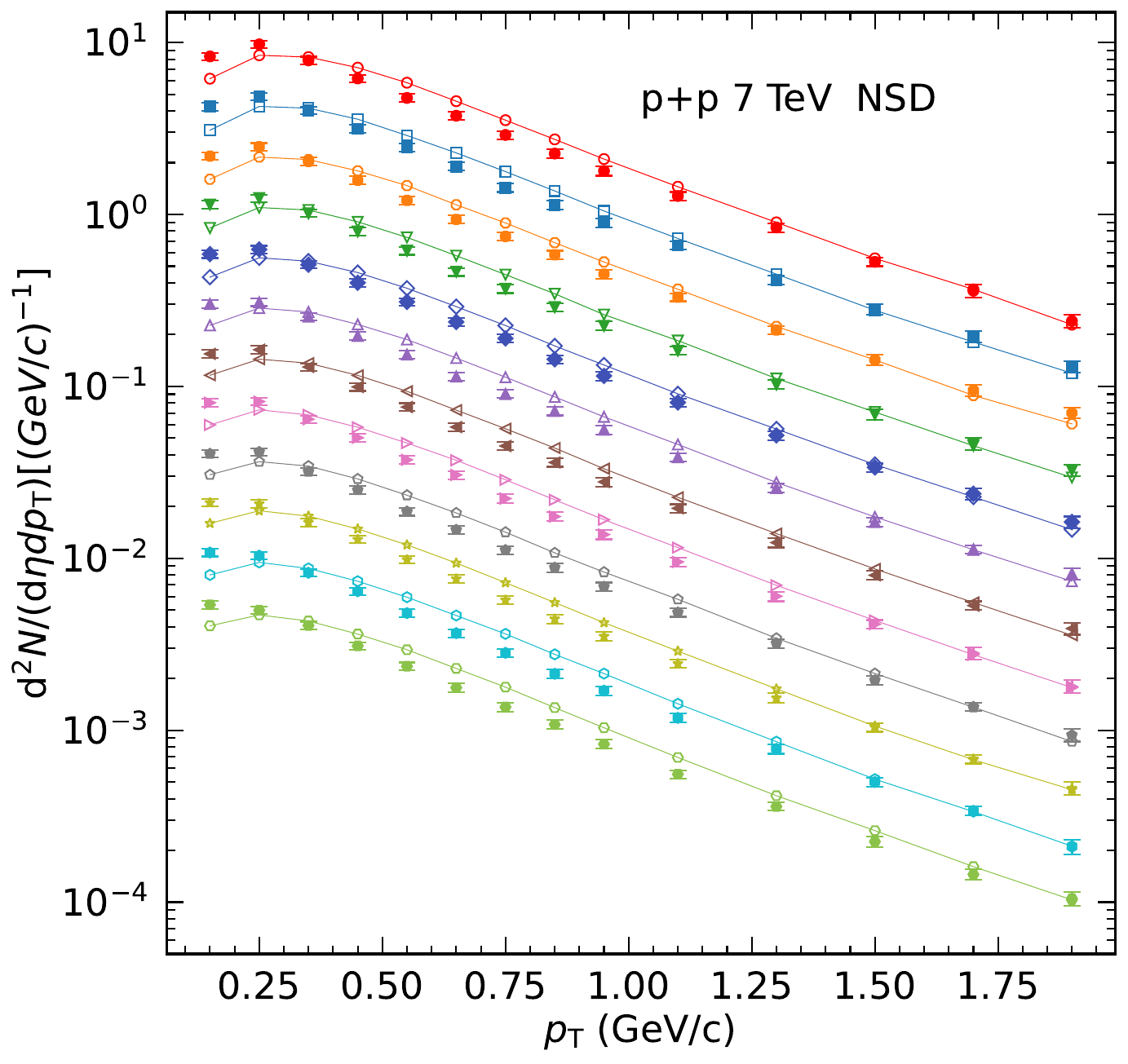}
    \end{overpic}
	}
   \caption{Transverse momentum spectra of charged particles in various pseudorapidity bins in NSD $pp$ collisions at $\sqrt{s}=$ 0.9, 2.36, and 7 TeV from PACIAE 4.0 model simulations (lines) are compared with CMS data (symbols) \cite{CMS:2010wcx,CMS:2010tjh}. For better visualization, a scaling factor indicated in the legend is applied to both simulation results and experimental data for each pseudorapidity bin.}
  \label{fig3}
\end{figure*}

Figure~\ref{fig2} presents the pseudorapidity density distributions of charged particles ($dN_{ch}/d\eta$) in NSD $pp$ collisions simulated by PACIAE 4.0 at $\sqrt{s}=$ 0.9, 2.36, and 7 TeV, compared with corresponding CMS experimental data. The experimental measurements are shown as black dots with error bars, while the simulation results are represented by red open triangles. The simulation reproduces the experimental data well, accurately capturing both the peak and valley positions of the distributions. Within the region $|\eta|<2$, $dN_{ch}/d\eta$ exhibits only weak pseudorapidity dependence, reflecting the characteristic uniformity of particle production at midrapidity, a well established feature observed in both $pp$ and $AA$ collisions. In the forward/backward regions $|\eta|>2$, $dN_{ch}/d\eta$ decreases rapidly with increasing $|\eta|$, which is attributed to kinematic and dynamical constraints on particle production at large rapidities. Both experimental data and simulations show the expected increase in total charged particle multiplicity with collision energy $\sqrt{s}$.

Figure \ref{fig3} shows the pseudorapidity dependence of the transverse momentum distributions of charged particles produced in NSD $pp$ collisions. Following experimental analysis procedures, the pseudorapidity space $|\eta|<2.4$ has been divided into 0.2-unit-wide bins for simulated NSD $pp$ collision events at $\sqrt{s}=$ 0.9, 2.36, and 7 TeV, respectively. For clearer visualization, both experimental data and the corresponding simulation data are scaled by factors indicated in the legend. The CMS data \cite{CMS:2010wcx,CMS:2010tjh} are shown as symbols, while PACIAE 4.0 simulation results appear as lines, with different colors distinguishing pseudorapidity bins. The model successfully reproduces transverse momentum distributions across all pseudorapidity bins simultaneously. At $p_{\rm{T}}<0.3$ GeV/c, PACIAE 4.0 underestimates the experimental data, while for $0.3 \leq p_{\rm{T}} \leq 1 $ GeV/c, the model slightly overpredicts the experimental data. Overall deviations are small and the maximum deviation for few points is within $40\%$, which is an excellent achievement for a transport model. Notably, PACIAE 4.0 matches the precision of Boltzmann-Gibbs blast wave model \cite{Waqas:2024suh} as well as EPOS-LHC and PYTHIA 8 \cite{Ajaz:2022mga}. We emphasize that both pseudorapidity density distributions and pseudorapidity dependence of the transverse momentum spectra of charged particles in NSD $pp$ collisions are studied simultaneously and systematically. Unlike INEL $pp$ collisions, where the value of parameter $b$ is adjusted for different collision energies \cite{Xie:2025vnh}, NSD collisions are successfully described using a single parameter set across all three collision energies studied, which is the ideal case for transport models.

\section{CONCLUSIONS}\label{con}

The newly released PACIAE 4.0 model is utilized to simulate NSD $pp$ collisions at $\sqrt{s}=$ 0.9, 2.36 and 7 TeV. The model demonstrates good agreement with CMS experimental data for both pseudorapidity density distributions and pseudorapidity dependence of the transverse momentum spectra, remarkably using a single parameter set across all three collision energies studied. Building on our previous work \cite{Xie:2025vnh} for INEL $pp$ collisions with more collision energies available, these results further validate the PACIAE 4.0 model as a robust tool for systematic studies of $pp$ collision physics. The model's reliability makes it particularly valuable for generating simulation data in experimentally unexplored regions. We note that reproducing experimental data requires distinct parameter sets for NSD and INEL $pp$ collisions at the same collision energy. This discrepancy may originate from either experimental analysis methodologies or inherent model design limitations in PACIAE. Addressing this issue will be the focus of future improvements. 

\section*{Acknowledgement}
 This work is supported by the National Natural Science Foundation of China under grant Nos. 11447024, 11505108 and 12375135, and by the 111 project of the foreign expert bureau of China. Y.L.Y. acknowledges the financial support from Key Laboratory of Quark and Lepton Physics in Central China Normal University under grant No. QLPL201805 and the Continuous Basic Scientific Research Project (No, WDJC-2019-13). W.C.Z. is supported by the Natural Science Basic Research Plan in Shaanxi Province of China (No. 2023-JCYB-012). H.Z. acknowledges the financial support from Key Laboratory of Quark and Lepton Physics in Central China Normal University under grant No. QLPL2024P01.

\bibliography{reference.bib}

\end{document}